\documentclass[prb,twocolumn,superscriptaddress,showpacs,showkeys,preprintnumbers,amsmath,amssymb,floatfix,aps,longbibliography]{revtex4-2}
\usepackage{graphicx}
\usepackage{bm}
\usepackage{placeins}
\usepackage{graphics}
\usepackage{color}
\usepackage{hyperref}
\usepackage{multirow}
\usepackage{blindtext}
\usepackage{bbold}
\usepackage{comment}
\begin{document}

\title{
Distinguishing nodal and nonunitary superconductivity in quasiparticle interference of an Ising superconductor with Rashba spin-orbit coupling: an example of NbSe$_2$
}

\author{Jozef Haniš}
    \email{jozef.hanis@student.upjs.sk}
	\affiliation{Institute of Physics, Pavol Jozef \v{S}af\'{a}rik University in Ko\v{s}ice, 04001 Ko\v{s}ice, Slovakia}
\author{Marko Milivojevi\'c}
	\affiliation{Institute of Informatics, Slovak Academy of Sciences, 84507 Bratislava, Slovakia}
        \affiliation{Faculty of Physics, University of Belgrade, 11001 Belgrade, Serbia}
\author{Martin Gmitra}
    \email{martin.gmitra@upjs.sk}
	\affiliation{Institute of Physics, Pavol Jozef \v{S}af\'{a}rik University in Ko\v{s}ice, 04001 Ko\v{s}ice, Slovakia}
	\affiliation{Institute of Experimental Physics, Slovak Academy of Sciences, Watsonova 47, 04001 Ko\v{s}ice, Slovakia}
\date{\today}
\begin{abstract}
The NbSe$_2$ monolayer with Rashba spin-orbit coupling represents a paradigmatic example of an Ising superconductor on a substrate.
Using a single-band model and symmetry analysis, we present general superconducting pairing functions beyond the nearest-neighbor approximation, uncovering new types of gap functions, including the nodal singlet gap function and the triplet non-unitary pairing function that breaks time-reversal symmetry. The non-unitarity builts in the asymmetrical band dispersion in the superconducting quasiparticle energy spectra. Performing exact T-matrix calculations of quasiparticle interference due to a single scalar impurity scattering, we found that the interference patterns possess characteristic features distinguishing the type of pairing and possible nematic and chiral symmetry violations. 
\end{abstract}
\maketitle
\section{Introduction}
Since its initial discovery~\cite{MLH10}, transition-metal dichalcogenides (TMDC) have attracted lots of attention due to their intriguing electronic structure \cite{KH12,CRG13} and potential application in many subfields of solid state physics such as spintronics \cite{ZCS11,KZD13,SYZ13,KGR13,ABX14,KBG15}, valleytronic \cite{LSS18}, optotronics \cite{PNK15}, and superconductivity \cite{RCG13,Hsu2017,MK18,OYK18,shaffer2020,HHS21,WHK21,KDL22,HLK22,Das2023:npjCM,roy2024:arXiv}.
TMDC crystalizing in several structural polytypes. Trigonal prismatic polytype is a stable structure of NbSe$_2$ exhibiting superconductivity in its bulk form \cite{Frindt1972PRL,Foner1973,khestanova2018,Noat2015} as well as down to a single layer limit \cite{Xi2016,delaBarrera2018}. Monolayer NbSe$_2$, unlike the bulk structure, lacks inversion symmetry, which leads to uniquely resolved large spin split bands due to spin-orbit coupling near the K points \cite{Xiao2021:PRL}. Due to time-reversal symmetry, the spins are locked to momentum with opposite directions in K and K$^\prime$ points and ${\bf D}_{3{\rm h}}$ symmetry, restricting the spins' orientation to the out-of-plane direction. Formed Cooper pairs break their rotational invariance in spin space leading to the novel pairing dubbed as Ising superconductivity \cite{Lu2015,Xi2016}. A key consequence of the Ising superconductivity is robustness to the in-plane magnetic fields exceeding considerably the Pauli limit \cite{Xi2016,delaBarrera2018}.
The Ising superconductivity is attracting significant attention \cite{Wickramaratne2023APL,xiong2024electrical, fang2024interplay, yang2024endowing, cohen2024josephson,patel2024electron} as it is considered as a principal mechanism for the giant in-plane upper critical magnetic field also observed in the bulk layered misfit structures \cite{Samuely2021PRB,Samuely2023PRB,bai2020multi}. 

Misfit layered compounds are a class of heterostructured materials that consist of two structurally different materials forming an ordered superstructure \cite{Meerschaut1996,Wiegers1996,Ng2022APLM}. 
(LaSe)$_{1.14}$(NbSe$_2$)$_x$, $x=1,2$ are examples in which the alternating LaSe and NbSe$_2$ layers form a slab stacked along their $c$ direction with a mismatch along the $a$ direction \cite{Roesky1993,Nader1998}. 
Electronic band structure near the Fermi level reminds Nb $d$-band with an offset due to strong electron doping, from 0.5 to 0.6 electron per Nb \cite{Leriche2020,Niedzielski2024:arxiv}.
The misfit structures, however, represent an unprecedented platform for charge transfer control \cite{Zullo2023:NL} providing access to topological superconductivity \cite{Hsu2017} due to the high tunability of chemical potential that can also be achieved by engineering La vacancies \cite{Samuely2023PRB} or alloying. 
The Ising superconductivity has been observed in the misfit structures \cite{Samuely2021PRB,bai2020multi,Leriche2020} suggesting that the spins near the K valleys still possess significant out-of-plane components.

Classification of the possible superconducting states according to the irreducible representations (IR) of a given symmetry group~\cite{FLS11,GFS12} can be performed without entering into the microscopical details of the origin of the superconductivity~\cite{TGW20,CG21,Horhold2023:2DM}.
Although there are some analyses of the superconducting TMDC systems with broken horizontal mirror plane symmetry~\cite{YML14}, the full classification of the possible pairing functions is still missing.
Doping control in the misfit structures \cite{Leriche2020}, prediction that superconductivity in gated MoS$_2$ \cite{Ye2021,Taniguchi2021} can possess exotic topological pairing \cite{YML14}, as well as the recent experiment~\cite{CLA22} reporting competition between the nodal and nematic superconductivity in a monolayer NbSe$_2$ on a substrate motivate us to study superconducting pairing in the reduced ${\bf C}_{\rm 3v}$ symmetry case. 
In such cases, both inversion and horizontal mirror plane symmetry are broken, giving rise to the Rashba spin-orbit coupling that tilts the spins to in-plane directions. 

Electronic states of TMDC materials can be well described near the K and $\Gamma$ valleys using a single-band effective tight-binding model \cite{SM19}. We extend the model considering spin-flip Rashba spin-orbit coupling term modeling effect of a substrate, gating or electron doping of the NbSe$_2$ monolayer in the reduced ${\bf C}_{\rm 3v}$ symmetry and constructed all possible superconducting pairing functions classified according to the irreducible representations (IRs)~\cite{VG85, K86,ABG99, KS19}. 

The benefit of the proposed general approach is that it allows us to go beyond the nearest-neighbor approximation for the pairing function. Consequently, we found novel exotic types of gap functions such as the nodal singlet and non-unitary triplet functions.

As quasiparticle interference (QPI) has been used to investigate superconducting gap function~\cite{AT13,HAE15,BSA20,Levitan2023:PRB,Rashid2023:SP}, we calculate QPI for each of the superconducting pairing functions. 
The obtained QPI patterns have characteristic features that allow distinguishing between the different types of pairing. 

This paper is organized as follows. In Section~\ref{Sec:Gaps} we study all possible types of a superconducting pairing function in a system with ${\bf C}_{3{\rm v}}$ symmetry.
In Section~\ref{Sec:BdG},  we present the numerical method for QPI calculation and define the effective quasiparticle BdG model for NbSe$_{2}$ using the single-band model of the NbSe$_2$ monolayer with Rashba spin-orbit coupling described in  Appendix~\ref{sec:DFT-TB}.

In Section~\ref{Results}, by assuming a single scalar impurity scattering, we analyze QPI patterns in the first Brillouin zone for different types of gaps. Final remarks are given in the Conclusions section, where we also summarize the main consequences of our results. 

\section{Gap functions classification according to irreducible representations}\label{Sec:Gaps}
\begin{figure}[t]
\centering
\includegraphics[width=6.5cm]{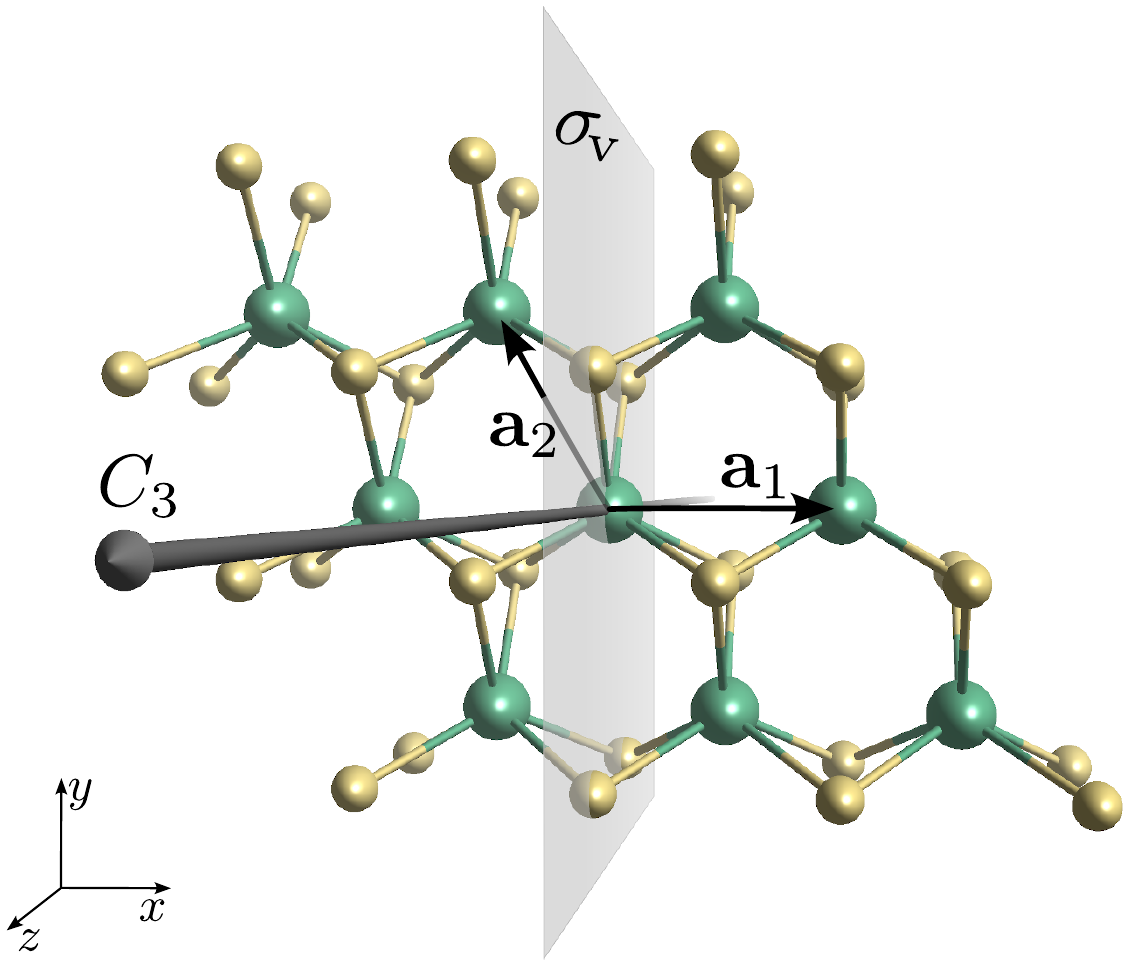}
\caption{Perspective view of the NbSe$_2$ monolayer crystal structure with the considered lattice vectors ${\bf a}_1$ and ${\bf a}_2$, vertical mirror plane $\sigma_{\rm v}$ that coincides with the $yz$-plane, and the main $C_3$ axis.}\label{fig1}
\end{figure}

In this Section, we will focus on the construction of the gap function for a system with the point group symmetry ${\bf C}_{3{\rm v}}$. The considered geometry of the system is shown in the Fig.~\ref{fig1}.
The group contains 6 elements and two generators: three-fold rotation $C_3$ and vertical mirror plane $\sigma_{\rm v}$.
The superconducting pairing can be transformed according to the one-dimensional (1D) irreducible representations (IR) $A_1$ or $A_2$, and two-dimensional (2D) IR $E$, see Table~\ref{tab:C3v}.
\begin{table}[t]
\setlength{\tabcolsep}{10pt}
\renewcommand{\arraystretch}{1.5}
\caption{Table of one-dimensional and two-dimensional IRs of the group ${\bm C}_{3{\rm v}}$, where $s=0,1,2$ represents different elements of the subgroup ${\bm C}_{3}$.}
\label{tab:C3v}
\centering
\begin{tabular}{c|cc}\hline
  IR  & $C_{3}^s$  & $\sigma_{\rm v} C_3^s$\\ \hline\hline
 $A_{1}$   & 1             & 1            \\ 
 $A_{2}$   & 1             & -1             \\
$E$ & $\begin{pmatrix}
   {\rm e}^{{\rm i}\frac{2\pi}3s} & 0 \\
    0 &{\rm e}^{-{\rm i}\frac{2\pi}3s} 
\end{pmatrix}$ &
$\begin{pmatrix}
   0 &  {\rm e}^{-{\rm i}\frac{2\pi}3s} \\
    {\rm e}^{{\rm i}\frac{2\pi}3s} & 0 
\end{pmatrix}$      \\ \hline\hline
\end{tabular}
\end{table}

Antisymmetry condition on the gap parameter due to the Pauli principle implies that in the single band case, the orbital part of the superconducting gap must be

symmetric/antisymmetric to the momentum change (${\bf k}\rightarrow{-\bf k}$) in the singlet/triplet case. The straightforward way to implement such a rule is to construct the order parameter using the even/odd function. Here, we will use the trigonometric $\cos/\sin$ function in such a way that the translation invariance of the system is preserved, with an argument that is a scalar function of the type ${\bf k}\cdot{\bf r}$.

The general formulation of the singlet and triplet gap functions can be written as follows
\begin{eqnarray}
\Delta_{\mu,m}^{\rm s}&=&\sum_{g\in {\bf C}_{3{\rm v}}}{\Gamma}_{mm}^{(\mu)*}(g)
\cos{\big({\bf k}\cdot\mathcal{D}^+(g^{-1}{\bf r})\big)}{\bf d}_0,\label{Eq.Ds}\\
\Delta_{\mu,m}^{{\rm t},z}&=&\sum_{g\in {\bf C}_{3{\rm v}}}\Gamma_{mm}^{(\mu)*}(g)
\sin{\big({\bf k}\cdot\mathcal{D}^+(g^{-1}{\bf r})\big)}\mathcal{D}^-(g^{-1}){\bf d}_z,\label{Eq.Dtz}\nonumber\\
\Delta_{\mu,m}^{{\rm t},xy}&=&\sum_{g\in {\bf C}_{3{\rm v}}}\Gamma_{mm}^{(\mu)*}(g)
\sin{\big({\bf k}\cdot\mathcal{D}^+(g^{-1}{\bf r})\big)}\mathcal{D}^-(g^{-1}){\bf d}_1,\nonumber\label{Eq.Dtxy}
\end{eqnarray}
where $\mu=A_1,A_2,E$ defines the IR, $m=1,\ldots,|\mu|$, where $|\mu|$ is the dimension of the IR $\mu$, and $\Gamma_{mm}^{(\mu)*}(g)$ is the conjugated matrix element of the IR $\mu$ for the given group element $g$.
The spin space is spanned conventionally by the matrices $d_l={\rm i}\sigma_l\sigma_y$ forming a pseudovector under rotation in spin space, where $\sigma_l$ are Pauli matrices for $l=x,y,z$, and $\sigma_0$ is the unit matrix. The component with $l=0$ corresponds to the antisymmetric part of the singlet pairing function, while $l=x,y,z$ components constitute the symmetric parts of the triplet pairing functions. 
Whereas the singlet gap function can be constructed using $d_0$ solely, in the case of the triplet pairing the gap function should be composed using the pseudovector
${\bf d}=(d_x,d_y,d_z)$. Due to the ${\bf C}_{3{\rm v}}$ symmetry, it is possible to decouple the $d_z$ component and the $(d_x,d_y)$ multiplet, leading to two types of triplet gap functions that can be constructed, see Eq.~\eqref{Eq.Ds}. This is the reason for choosing the ${\bf d}_z$ and ${\bf d}_1$ vectors, being equal to $(0,0,d_z)$ and $(d_x,0,0)$, and belonging to the orthogonal subspaces of the pseudovector space.

The components of the normal vector and pseudovector transform according to the matrix representations equal to
\begin{eqnarray*}
\mathcal{D}^\pm(C_3)=\begin{pmatrix}
    \cos{\frac{2\pi}3} & -\sin{\frac{2\pi}3} & 0  \\
    \sin{\frac{2\pi}3} & \cos{\frac{2\pi}3} & 0   \\
      0 & 0 & 1\end{pmatrix},
\mathcal{D}^\pm(\sigma_{\rm v})=\pm\begin{pmatrix}
    -1 & 0 & 0  \\
    0& 1 & 0   \\
    0 & 0 & 1
\end{pmatrix}.
\end{eqnarray*}
Representation for the other group elements can be obtained using the multiplication rule $\mathcal{D}(g_1)\mathcal{D}(g_2)=\mathcal{D}(g_1g_2)$. 

\subsection{Singlet gaps}
Here we analyze the contribution of the first neighbors and start the construction of the gap functions using ${\bf r}_1=a {\bf e}_x$ as the initial coordinate that coincides with one of the Bravais lattice vectors (${\bf a}_1 = {\bf r}_1$ and ${\bf a}_2 = {\bf r}_2$). 
The other vectors are obtained by using the action of the elements of the ${\bf C}_{3{\rm v}}$ group:
${\bf r}_2=\mathcal{D}^+(C_3){\bf r}_1$,
${\bf r}_3=\mathcal{D}^+(C_3^2){\bf r}_1$,
${\bf r}_4=\mathcal{D}^+(\sigma_{\rm v}){\bf r}_1$,
${\bf r}_5=\mathcal{D}^+(\sigma_{\rm v}C_3){\bf r}_1$,
${\bf r}_6=\mathcal{D}^+(\sigma_{\rm v}C_3^2){\bf r}_1$.
The $k$-dependent functions that can be used to construct the gap function are $v_i={\bf k}\cdot{\bf r}_i$,  $(i=1,\ldots,6)$. Note that under an element $g$ of the point group, ${\bf k}\rightarrow g{\bf k}$ and $f({\bf k}\cdot{\bf r})$ is mapped into $f((g{\bf k})\cdot{\bf r})=f({\bf k}\cdot(g^{-1}{\bf r}))$.

Using the symmetrization procedure in Eq.~(\ref{Eq.Ds}), we can construct the order parameter function with $\cos$ having $v_i$ (or some linear combination) as an argument. The gap
function in the IR $A_1$ has the form
\begin{equation}\label{SingletAo}
\Delta_{A_1}^{\rm s}=\Big[2\cos{(k_x a)}+4\cos{\frac{k_x a}2}\cos{\frac{\sqrt{3}k_y a}2}\Big]d_0.
\end{equation}
Note that, around the $\Gamma$ point,
$\Delta_{A_1}^{\rm s}\approx d_0$, resembling the well known $s$-wave gap. 

The IR $E$ is a two-dimensional representation for which we construct the order parameter in the two-component form $\Delta_E^{\rm s}=(\Delta_{E,1}^{\rm s},\Delta_{E,2}^{\rm s})$. The components are connected by the vertical mirror plane symmetry $\sigma_{\rm v}$ as $\sigma_{\rm v} \Delta_{E,1}^{\rm s}=\Delta_{E,2}^{\rm s}$ and read
\begin{eqnarray}\label{SingletE1}
\Delta_{E,1/2}^{\rm s}
&=&\Big[\cos{(k_xa)}-\cos{\frac{k_xa}{2}}\cos{\frac{\sqrt{3}k_ya}{2}}\nonumber\\
&&\pm{\rm i}\sqrt{3}\sin{\frac{k_xa}{2}}\sin{\frac{\sqrt{3}k_ya}{2}}
\Big] d_0.
\end{eqnarray}
Close to the $\Gamma$ point, $\Delta_{E,1/2}^{\rm s}\approx (k_x^2-k_y^2)\pm {\rm i} k_xk_y$,
resembling the $d+\mathrm{i}d$ type of a superconducting gap. 

In the first neighbor approximation, the order parameter that transforms according to the IR $A_2$ is zero. To obtain a nonzero gap function one needs to consider for the symmetrization procedure the initial vector 
${\bf r} \rightarrow 2{\bf a}_1-{\bf a}_2$ 
which corresponds to the third neighbor. The gap function then reads
\begin{eqnarray}\label{SingletBo}
\Delta_{A_2}^{\rm s}
&=&2\Big[
\sin{\frac{k_xa}2}\sin{\frac{3\sqrt{3}k_ya}2}-
\sin{(2k_xa)}\nonumber\\
&&\times\sin{(\sqrt{3}k_ya)}+\sin{\frac{5k_xa}2}\sin{\frac{\sqrt{3}k_ya}2}\Big] d_0.\nonumber
\end{eqnarray}
The leading polynomial around the $\Gamma$ is of the sixth order and equals $k_x^5k_y-10/3 k_x^3k_y^3+k_xk_y^5$. If we express this term in polar coordinates, we find that $\Delta_{A_2}^{\rm s}\approx \sin{(6\varphi)}d_0$, has six nodal lines.
\subsection{Triplet gaps}
In the case of the triplet pairing, the order parameter is composed using the pseudovector ${\bf d}=(d_x,d_y,d_z)$, with components in the matrix form equal to
\begin{equation}
d_x=\begin{pmatrix}
   -1 & 0 \\
    0 & 1  \\
\end{pmatrix},
d_y=\begin{pmatrix}
   {\rm i} & 0 \\
    0 & {\rm i}  \\
\end{pmatrix},
d_z=\begin{pmatrix}
    0 & 1 \\
    1 & 0  \\
\end{pmatrix}.
\end{equation}
To construct an irreducible pseudovector, we will use the $\sin$ functions (since the gap must be antisymmetric under the change ${\bf k}\rightarrow{-\bf k}$), while we can independently use the multiplet of $(d_x,d_y)$ that transform according to the IR $E$ and the pseudocomponent $d_z$, transforming according to the representation $A_2$. Due to their different symmetry properties, we construct the gap function using independently $d_{z}$ and the $(d_x,d_y)$ multiplet.

\subsubsection{Triplets gap functions with $d_z$ pseudovector component}\label{sec:triplet_dz_gaps}
To construct triplet gap functions for the $d_z$ pseudovector component solely, i.e., ${\bf d}_z=(0,0,d_z)$, we consider for the initial coordinate the vector ${\bf r}={\bf a}_1$, meaning that the nearest-neighbor approximation is employed. The gap function for the $A_1$ representations equals
\begin{eqnarray}\label{Aotriplet}
\Delta_{A_1}^{{\rm t},z}&=&\Big[\Big(\cos{\frac{k_xa}{2}}-\cos{\frac{\sqrt{3}k_y a}{2}}\Big)\sin{\frac{k_xa}2}\Big]d_z.
\end{eqnarray}
Around the $\Gamma$ point, this gap function
can be approximated as $k_x(3k_y^2-k_x^2)d_z$, representing the $f$-wave gap.

In the case of the $A_2$ IR, we are unable to obtain the gap function considering the nearest neighbors, but in the case of second-neighbors, for the initial coordinate ${\bf r}\to{\bf a}_1-{\bf a}_2$, the nonzero gap of the form
\begin{eqnarray}\label{Botripletdz}
\Delta_{A_2}^{{\rm t},z}=\Big[\sin{(\sqrt{3}k_ya)}-2\sin{\frac{\sqrt{3}k_ya}{2}}
\cos{\frac{3k_xa}{2}}\Big]d_z
\end{eqnarray}
is constructed. Around the $\Gamma$ point, the $\Delta_{A_2}^{{\rm t},z}$ gap function
can be approximated as $k_y(k_y^2-3k_x^2)d_z$, suggesting the $f$-wave character of superconducting order parameter.

For the gap function within the two-dimensional IR $E$ considering the first neighbors we obtain
\begin{eqnarray}\label{E1tripletdz}
\Delta_{E,1/2}^{{\rm t},z}
&=&\Big[
\sin{(k_x a)}+\sin{\frac{k_xa}2}\cos{\frac{\sqrt{3}k_y a}2}\nonumber\\
&&\pm{\rm i}\sqrt{3}\cos{\frac{k_xa}2
\sin{\frac{\sqrt{3}k_ya}2}}\Big] d_z.
\end{eqnarray}
We have additionally checked that the relation $\sigma_{\rm v}\Delta_{E,1}^{{\rm t},z}=\Delta_{E,2}^{{\rm t},z}$ is satisfied, meaning that the constructed gap fulfills the required symmetry.
In the vicinity of the zone center, the gap can be approximated as $(k_x\pm {\rm i}k_y)d_z$, resembling the $p+\mathrm{i}p$ gap.

\subsubsection{Triplets gap functions with $d_x$ and $d_y$ pseudovector components} \label{sec:triplet_dx_dy_gaps}
We now construct triplet gaps using the doublet $(d_x,d_y)$. The gap function transforming according to the IR $A_1$ is zero for nearest neighbor coordinates. The non-zero contribution we get for ${\bf r} \to {\bf a}_1+{\bf a}_2$ reading
\begin{eqnarray}\label{AotripletXY}
\Delta_{A_1}^{{\rm t},xy}
&=&\Big[ 3\cos{\frac{k_x a}2} \sin{\frac{\sqrt{3} k_ya}2}\Big] d_x-\Big[\sqrt{3}\Big(2\cos{\frac{k_xa}2}\nonumber\\
&&+\cos{\frac{\sqrt{3}k_y a}2}\Big)\sin{\frac{k_xa}2}\Big] d_y.
\end{eqnarray}

Around the $\Gamma$ point, the gap function can be approximated as
$k_yd_x-k_xd_y$, consistent with the previous results based on the method of invariants~\cite{SAS17}.

In the case of the IR $A_2$ representation, we can obtain a nonzero gap function considering nearest-neighbors in the form 
\begin{eqnarray}\label{BotripletXY}
\Delta_{A_2}^{{\rm t},xy}
&=&\Big[\Big(2\cos{\frac{k_xa}2}+\cos{\frac{\sqrt{3}k_y a}2}\Big)\sin{\frac{k_xa}2}\Big] d_x\nonumber\\
&&+\Big[\sqrt{3}\cos{\frac{k_x a}2} \sin{\frac{\sqrt{3} k_ya}2}\Big] d_y.
\end{eqnarray}
Around the $\Gamma$ point, the gap function $\Delta_{A_2}^{{\rm t},xy}$ can be approximated as $k_xd_x+k_yd_y$, as shown in~\cite{SAS17}.
Finally, for the two-dimensional IR $E$
we were unable to construct the gap function using the initial coordinate ${\bf r}=a {\bf e}_x$ for nearest-neighbors.

The relation $\sigma_{\rm v}\Delta_{E,1}^{{\rm t},xy}=\Delta_{E,2}^{{\rm t},xy}$ is not satisfied, but rather $\sigma_{\rm v}\Delta_{E,1}^{{\rm t},xy}=-\Delta_{E,2}^{{\rm t},xy}$.
The solution  $\Delta_{E}^{{\rm t},xy}=(\Delta_{E,1}^{{\rm t},xy}, \Delta_{E,2}^{{\rm t},xy})$ that has the correct symmetry properties with respect to the vertical mirror symmetry can be constructed using the initial vector ${\bf r} \to {\bf a}_1+2{\bf a}_2$ 
\begin{eqnarray}\label{tripletE1xy}
\Delta_{E,1/2}^{{\rm t},xy}
&=&\Big[2\sin{(\sqrt{3}k_y a)}-\cos{\frac{3k_x a}2}\sin{\frac{\sqrt{3}k_ya}2}\nonumber\\
&&\pm{\rm i}\sqrt{3}\sin{\frac{3k_x a}2}\cos{\frac{\sqrt{3}k_y a}{2}}\Big] d_x\nonumber\\
&&+\Big[\sqrt{3}\sin{\frac{3k_xa}2}\cos{\frac{\sqrt{3}k_ya}2}\nonumber\\
&&\mp3{\rm i}\cos{\frac{3k_x a}2}\sin{\frac{\sqrt{3}k_y a}2}\Big] d_y.
\end{eqnarray}
Around the $\Gamma$ point, the gap function can be approximated as
$\Delta_{E,1/2}^{{\rm t},xy}\approx (k_xd_y+k_yd_x)\pm{\rm i}(k_xd_x-k_yd_y)$.
Furthermore, it can be shown that the gap functions $\Delta_{E,1/2}^{{\rm t},xy}$ are non-unitary. According to~\cite{R22}, the unitary gap function $\Delta({\bf k})$ satisfies the equation
\begin{equation}\label{unitaritycindition}
    \Delta({\bf k})\Delta^{\dag}({\bf k})=\Delta_U^2({\bf k})\mathcal{I}_2,
\end{equation}
where $\Delta_U^2({\bf k})$ is the norm of the unitary gap, while $\mathcal{I}_2$ is the identity $2\times 2$ matrix. By analyzing the equation
\begin{equation}
\Delta_{E,1/2}^{{\rm t},xy}(\Delta_{E,1/2}^{{\rm t},xy})^{\dag}=4\begin{pmatrix}
A^2({\bf k}) & 0 \\
0 & B^2({\bf k})
\end{pmatrix},
\end{equation}
where
\begin{eqnarray}
    A^2({\bf k})&=&\Big[-2\cos{\frac{3k_x a}2}\sin{\frac{\sqrt{3}k_ya}2}+\sin{(\sqrt{3}k_y a)}\Big]^2,\nonumber\\
    B^2({\bf k})&=&3\sin^2{(\frac{3k_x a}2)}\cos^2{(\frac{\sqrt{3}k_ya}2)}+\big(\cos{\frac{3k_xa}2}\nonumber\\
&&\times\sin{\frac{\sqrt{3}k_ya}2}+\sin{(\sqrt{3}k_y a)}\big)^2,
\end{eqnarray}
one could conclude that the gap functions $\Delta_{E,1/2}^{{\rm t},xy}$ are non-unitary, since the functions $A^2({\bf k})$ and  $B^2({\bf k})$ are not identically equal. As an illustration, we expand the functions $A^2({\bf k})$ and $B^2({\bf k})$ around the $\Gamma$ point up to the ${\bf k}^2$ term and get $A^2({\bf k})\approx 0$, 
$B^2({\bf k})\approx 27/4 (k_x^2+k_y^2)a^2$.
We note that the gap functions $\Delta_{E,1}^{{\rm t},xy}$ and $\Delta_{E,2}^{{\rm t},xy}$ could realize a certain linear combination $c_1\Delta_{E,1}^{{\rm t},xy}+c_2\Delta_{E,2}^{{\rm t},xy}$. Assuming that $c_1$ and $c_2$ are arbitrary complex numbers, the unitarity given by Eq.~\eqref{unitaritycindition} is restored if and only if the relation $|c_1|^2=|c_2|^2$ is satisfied.

\section{Effective quasiparticle {BdG} model for {NbSe$_{2}$}}\label{Sec:BdG}

We are interested in studying the effects of electron superconducting pairing in monolayer NbSe$_2$ at a low-temperature regime,  neglecting superconducting phase fluctuation \cite{Leggett1966}. We model the superconductivity by Bogoliubov de-Gennes (BdG) formalism with Hamiltonian 
\begin{equation}
    \mathcal{H}_\mathrm{BdG} = \frac{1}{2} \sum_\mathbf{k} \Psi^\dagger_\mathbf{k} \mathcal{H}_{\mathbf{k}}^\mathrm{BdG} \Psi_\mathbf{k},
\end{equation}
considering the Nambu spinor, $\Psi^\dagger_{\mathbf{k}} = [ c^\dagger_{\mathbf{k}\uparrow}, c^\dagger_{\mathbf{k}\downarrow}, c_{\mathbf{-k}\uparrow}, c_{\mathbf{-k}\downarrow} ]$ with fermionic creation $c^\dagger_{\mathbf{k}\uparrow}$ and annihilation $c_{-\mathbf{k}\uparrow}$ operators. The $4\times 4$ BdG Hamiltonian in reciprocal space $\mathcal{H}_{\mathbf{k}}^{\mathrm{BdG}}$ reads~\cite{LS18}
\begin{equation}
    \mathcal{H}_{\mathbf{k}}^{\mathrm{BdG}} =
    \begin{pmatrix}
        H_\mathrm{e}(\mathbf{k}) & \Delta(\mathbf{k})\\
        \Delta^{\dag}(\mathbf{k})& -H_\mathrm{e}^{\mathrm T}(-\mathbf{k})
    \end{pmatrix},
\end{equation}
where the electron-like Hamiltonian $H_\mathrm{e}(\mathbf{k})$ is the effective tight-binding Hamiltonian 
\begin{equation}
    H_\mathrm{e}(\mathbf{k}) =  \mathcal{H}_\mathrm{orb}(\mathbf{k}) + \mathcal{H}_\mathrm{I}(\mathbf{k}) + \mathcal{H}_\mathrm{R}(\mathbf{k}),
\end{equation}
with the orbital part $\mathcal{H}_\mathrm{orb}$ describing the dispersion of the Nb $d$-band in the vicinity of the Fermi level, intrinsic $\mathcal{H}_\mathrm{I}(\mathbf{k})$ and Rashba spin-orbit coupling $\mathcal{H}_\mathrm{R}(\mathbf{k})$ contributions. More details of the Hamiltonian are discussed in Appendix \ref{sec:DFT-TB}.

The superconducting order parameter is equal to $\Delta_{\mathbf{k}} = \Delta_0 \Delta^\mathcal{P}_r(\mathbf{k})$, where $\Delta_0$ is the gap amplitude, while $\Delta^\mathcal{P}_r(\mathbf{k})$ is the $\mathbf{k}$-dependent part of the gap function respecting the point group symmetry representations, $r = \{A_1, A_2, E_{1/2}\}$, for singlet and triplet pairings, $\mathcal{P} = \{ {\rm s}, {\rm t} \}$, as discussed in Section \ref{Sec:Gaps}. 
Superconducting critical temperature of monolayer 1H-NbSe$_2$ is $T_c \approx 2~\mathrm{K}$ \cite{ugeda2016, wang2017, khestanova2018}. This yields from BCS theory for superconducting gap $\Delta(0~\mathrm{K})\approx 0.3$~meV. For simplicity, in what follows, we will use $\Delta_0=1$ meV, approaching the same order of magnitude as the extracted superconducting gap.

\subsection{QPI \& Spectral functions}\label{QPI}
QPI is extracted from scanning tunneling spectroscopy measurements by examining the Fourier transform of $\mathrm{d}I/\mathrm{d}V$ maps of the local density of states (LDOS).
It is a quite powerful qualitative tool and represents a unique probe of wavelengths of LDOS-modulated oscillations caused by impurities present in the system, which in turn contains information on the electronic structure of the pure metallic \cite{Crommie1993,sprunger1997}, semiconducting \cite{Wittneven1998,Kanisawa2001} and superconducting \cite{Byers1993,hoffman2002} systems. 
Calculated QPI patterns are similar to those experimentally observed \cite{Wang2003PRB} and have been used for identification of the disorder type \cite{balatsky2006}, superconducting gap features such as phase structure, sign \cite{Pereg-Barnea2003,Nummer2006} and orbital order \cite{Chen2023NatComm}.
It has been shown that quasiparticle responses of many impurities possess identical poles as a single impurity case \cite{capriotti2003,zhu2004}, therefore, we approximate the description of the elastic scattering process on a spin-conserving single impurity. Such impurity potential can be written in real Nambu space as
\begin{equation}
V = v \tau_3 \otimes \sigma_0,
\end{equation}
where $v$ is the impurity strength, $\tau_3$ is the Pauli matrix for electron-holes and $\sigma_0$ is the identity operator in the spin space. Considering the potential $V$ as a perturbation, the LDOS can be calculated using the T-matrix approach via Green's function written as
\begin{equation}
G(\mathbf{k},\mathbf{k}'; \omega) = G_0(\mathbf{k},\mathbf{k}'; \omega)+ \delta G(\mathbf{k},\mathbf{k}'; \omega)
\label{eq_GF}
\end{equation}
with
\begin{equation}
\delta G(\mathbf{k},\mathbf{k}'; \omega)=
\sum_{\mathbf{k}_1,\mathbf{k}_2}G_0(\mathbf{k},\mathbf{k}_1; \omega) T(\mathbf{k}_1, \mathbf{k}_2;\omega ) G_0(\mathbf{k}_2, \mathbf{k}' ; \omega),
\end{equation}
where $G_0(\mathbf{k},\mathbf{k}'; \omega)$ is retarded bare Green's function and $T(\mathbf{k}, \mathbf{k}'; \omega)$ is the T-matrix, represent the solution of the scattering problem for single scalar impurity \cite{economou2006}, and $\omega = \epsilon + \mathrm{i}\delta$ is the quasiparticle energy with small lifetime broadening ($\delta\approx 0.1$~meV).
The spin-conserving scalar impurity $V$ leads to a T-matrix independent of momentum. Hence, the T-matrix reads 
\begin{equation}
T(\omega) = V \cdot [\mathbb{1} - V \cdot G_0(\omega)]^{-1}
\end{equation}
where $G_0(\omega) = 1/\Omega \int [ \omega - \mathcal{H}_0(\mathbf{k})]^{-1}$ is the integrated bare Green's function within the first Brillouin zone (BZ). We assume in normal phase $\mathcal{H}_0(\mathbf{k}) = H_{\rm e}(\mathbf{k}),$ and for superconducting phase $\mathcal{H}_0(\mathbf{k}) = \mathcal{H}_{\mathbf{k}}^{\rm BdG}.$ 
As QPI describes LDOS of scattered quasiparticle process with momentum $\mathbf{k} \rightarrow \mathbf{k} + \mathbf{q}$, we substitute momentum $\mathbf{k}'$ with $\mathbf{k}' = \mathbf{k} + \mathbf{q}$ and integrate the Green's function (\ref{eq_GF}) over all possible $\mathbf{k}$ points in the first BZ. 
After the integration, the scattered quasiparticle amplitude can be expressed as follows 
\begin{equation}
	\rho(\mathbf{q}; \omega) = -\frac{1}{\pi} \mathrm{Im} \left\{ \frac{1}{\Omega} \int \mathrm{d}\mathbf{k}\, \delta G(\mathbf{k}, \mathbf{k} + \mathbf{q}; \omega) \right\}.
\end{equation}
The QPI $\rho(\mathbf{q}; \omega)$ can be evaluated via convolution theorem \cite{kohsaka2017}, which reduces computational complexity allowing to study of fine grid maps with small broadening.

For calculating the spectral characteristics of the band structure we use an imaginary part of the spectral function defined as
\begin{equation}
    \mathcal{A}(\mathbf{k}; \omega) = -\frac{1}{\pi} \mathrm{Im} \{ \mathrm{Tr} [ G_0(\mathbf{k};\omega) \cdot ( \mathbb{1} - V \cdot G_0(\omega) )^{-1} ] \},
\end{equation}
obtained from Eq.~(\ref{eq_GF}) taking $\mathbf{k}' = \mathbf{k}$. The Bloch spectral (density) function has interpretation as $\mathbf{k}$-resolved DOS \cite{Ebert2011:RPP}.

\section{QPI patterns}\label{Results}
In this section, we discuss calculated QPI maps separately for each derived superconducting gap function $\Delta_r^p(\mathbf{k})$ within the $\mathbf{C}_{3{\rm v}}$ symmetry. The QPIs were calculated for the Fermi energy of the single-band electron Hamiltonian $H_{\rm e}(\mathbf{k})$, see Appendix \ref{sec:DFT-TB}, that effectively describes doped NbSe$_2$ by 0.56 electrons. Such doping corresponds to a rigid shift of the Nb $d$-band as reported for (LaSe)$_{1.14}$(NbSe$_2$)$_2$ misfit using angle-resolved photoemission spectroscopy, scanning tunneling microscopy, and quasiparticle interference measurements \cite{Leriche2020}.
Similar electron doping was reported also for intercalated bulk NbSe$_2$ by imidazole cations \cite{Zhang22}.

\begin{figure}[t]
    \centering
    \includegraphics[width=7.0cm]{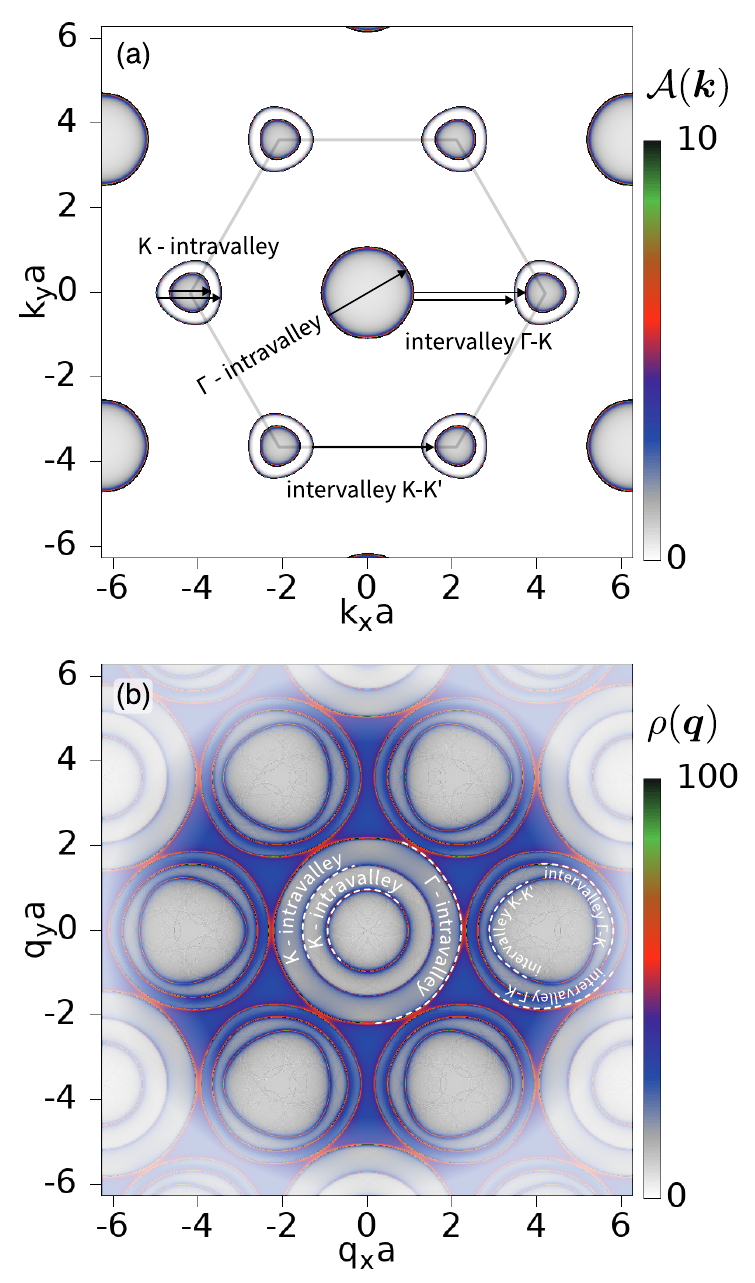}
    \caption{Calculated spectral characteristics for doped NbSe$_2$ monolayer by 0.56 electrons in normal state.
    (a)~spectral function $\mathcal{A}(\mathbf{k})$ at the Fermi level;
    (b)~QPI map for the scalar impurity with the potential $v=-0.1$~eV.}
    \label{fig:single_particle}
\end{figure}

In Fig.~\ref{fig:single_particle}(a) we show spectral function $\mathcal{A}(\mathbf{k})$ for the doped NbSe$_2$ monolayer at the Fermi level with the Fermi pockets around the $\Gamma$ and K points. The QPI map for the scalar impurity with the potential of $v=-0.1$~eV is shown in Fig.~\ref{fig:single_particle}(b). For small $\mathbf{q}$ momenta centered around the $\Gamma$ point there are three distinct contrasts forming contours that correspond to $\Gamma$-intravalley and two K-intravalley scattering processes, see schematically sketched $\mathbf{q}$ vectors in the Fig.~\ref{fig:single_particle}(a). The two distinct K-intravalley contours reflect strong spin-orbit split bands at the K pockets. For large scattering $\mathbf{q}$ we identified K-point centered patterns: intervalley ${\rm K-K'}$ scattering, and two contours corresponding to the $\Gamma - {\rm K}$ intervalley scattering implying the effect of the spin-orbit split bands around the K point.

\begin{figure*}[t]
\centering
    \includegraphics[width=16cm]{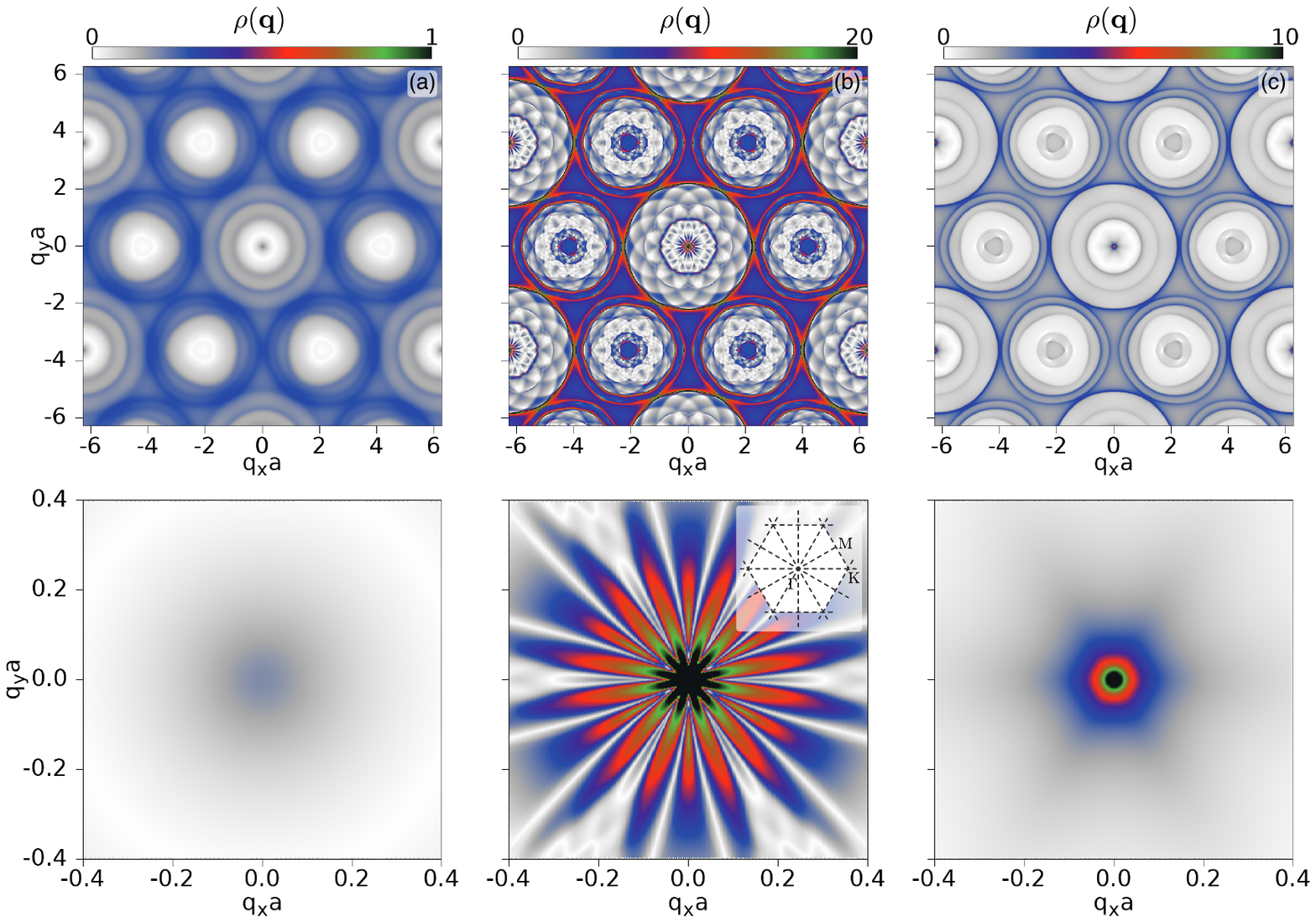} 
    \caption{Calculated QPI patterns in the superconducting state for singlet order parameters for (a)~$A_1$, (b)~$A_2$, and (c)~$E$ representations. Plots in the bottom row are the zone center zooms for the corresponding top row. The inset shows the first Brillouin zone with nodal lines. For calculations superconducting pairing amplitude $\Delta_0 = 1$~meV, the real part of the quasiparticle energy $\omega$ equal to Fermi energy, and the amplitude of the scalar impurity potential $v = -0.1$~meV was used.
}
\label{fig:QPI_singlet}
\end{figure*}

Bogoliubov quasiparticles can be viewed as Bloch states with modulated dispersion for which an elastic scattering can also result in modulation of the QPI signal. 
As the single-particle bandstructure is known in our case, from the contours at the Fermi energy, see Fig.~\ref{fig:single_particle}, $\mathbf{k}$-dependence of the gap function, and density of states (DOS), see Appendix~\ref{sec:DOS}, we can identify the origin of the most relevant scattering channels.
We start by examining QPI map for the singlet gap functions. 
In Fig.~\ref{fig:QPI_singlet}(a) we show a calculated QPI map for $A_1$ representation. The finite smearing of the entirely gaped system, see Fig.~\ref{fig:DOS}(a), leads to a QPI signal that preserves contour features discussed for the normal case. 

For $A_2$ representation the $\Gamma$-intravalley and intervalley $\Gamma-$K scatterings maintain significant contributions. 
The nodal lines connecting $\Gamma-$K, $\Gamma-$M, and K$-$K' points lead to a rich QPI pattern resembling a dahlia flower, see Fig.~\ref{fig:QPI_singlet}(b). The in-gap states for the $\Delta^{\rm s}_{A_2}$ gap function, see the v-shaped DOS dependence shown in Fig.~\ref{fig:DOS}(b), lead to the small $q$ momenta scattering events well-visible near the $\Gamma$ point as an inner corolla. It combines the small momentum scattering near the $\Gamma$ and K pockets resulting in 18 petals. 12 petals point to regions in between the nodal lines, and six petals point along the $\Gamma-$M lines that contain the $\sigma_{\rm v}$ plane. The dahlia-like QPI pattern near the $\Gamma$ point follows the six-fold symmetry of the $\Delta^{\rm s}_{A_2} \approx \sin(6\varphi) d_0$, see the zone center zoom in Fig.~\ref{fig:QPI_singlet}(b).

In the case of $E$ representations, for the 2D multiplet $(\Delta_{E,1},\Delta_{E,2})$ we will restrict our analysis to the first component $\Delta_{E,1}$ only, representing the energetically most favorable situation~\cite{ABG99}.
The QPI pattern for the singlet pairing shows similar contours as the normal phase. The gap function $\Delta_{E,1}^{\rm s}$ opens a global gap at the Fermi level, see Fig.~\ref{fig:DOS}(c). For the $\Gamma$ pocket, the gap function amplitude is smaller than for the K pockets. This leads within the finite broadening to the enhanced $\Gamma$ intravalley contour, and $\Gamma -$K intervalley contours. The finite broadening and reduced gap amplitude is also responsible for the enhanced signal at the zone center.

\begin{figure*}[t]
\centering
    \includegraphics[width=16cm]{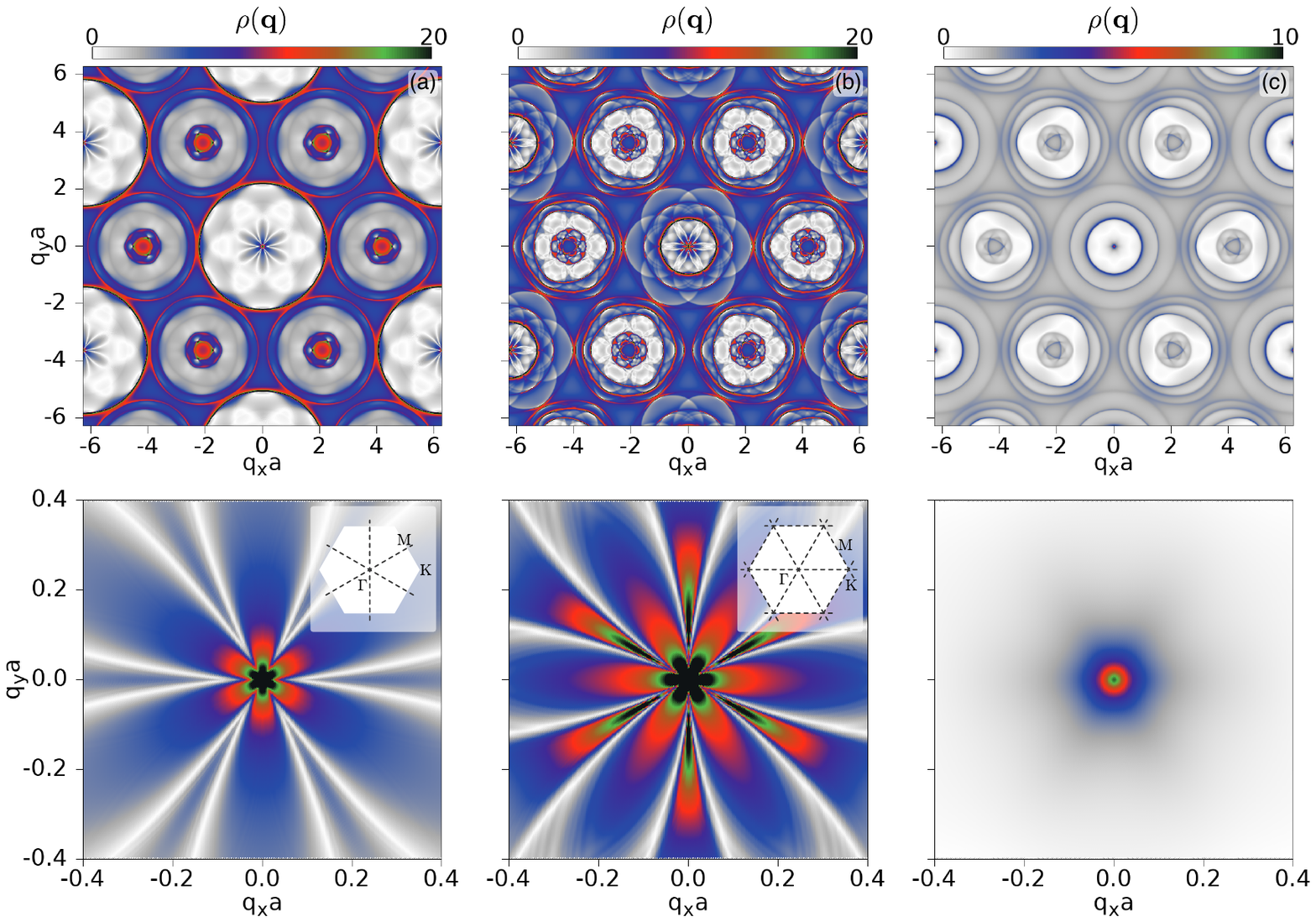} 
\caption{Calculated QPI patterns in the superconducting state for triplet order parameters with pseudovector component $d_{z}$ for (a)~$A_1$, (b)~$A_2$, and (c)~$E$ representations. Plots in the bottom row correspond to the zone center zoom. The insets show the first BZ with corresponding nodal lines. The other parameters are as in Fig.~\ref{fig:QPI_singlet}.
}
\label{fig:QPI_triplet_z}
\end{figure*}

We now move on to the case of the triplet pairing with $d_z$ pseudovector component. 
In Fig.~\ref{fig:QPI_triplet_z} we plot QPI map for $A_1$, $A_2$ and $E$ representations. 
In the case of the $A_1$ representation the K-intravalley and K$-$K' intervalley scatterings are suppressed. Instead, we observe a significant signal enhancement close to the $\Gamma$ and K points, see Fig.~\ref{fig:QPI_triplet_z}(a).
The enhancement is related to the presence of the nodal lines connecting $\Gamma -$M points. 
The $\Gamma$ pocket serves 6-fold symmetric in-gap states.
As the quasiparticles can scatter preferentially along the nodal lines with small momentum due to finite lifetime, the QPI signal forms a daisy-like flower pattern with six petals along the nodal lines, see the zoom near the zone center and inset of the first BZ showing the nodal lines. 
The fine structure around the K point has 3-fold symmetry and the sharp QPI signal is governed by the umklapp processed combining quasiparticles scattered off the nodal line in-gap states within the $\Gamma$ pocket and the states within the K pocket. The K pocket states are gaped, residing within the coherent peaks, but due to a finite broadening they offer a significant number of states to absorb the scattering.
The gap function for the $A_2$ representation possesses six-fold nodal symmetry at the $\Gamma$ and K points. The nodal lines connect $\Gamma -$K and K$-$K' points. The nodal quasiparticles modulate the QPI map such that it resembles a daisy flower pattern, see Fig.~\ref{fig:QPI_triplet_z}(b). The pattern is rotated by 30 degrees compared to the $A_1$ representation. The nodal character of the gap function $\Delta_{A_2}^{{\rm t},z}$ enhances the QPI signal on the apexes of the petals pointing along the nodal lines. The nodal character of the gap function also contributes to the small $\mathbf{q}$ momenta scattering forming characteristic inner corolla near the $\Gamma$ point. Close to the K point the 3-fold inner pattern is rotated by 60 degrees compared to $A_1$ representation signaling the contribution from the K$-$K' scatterings. We note that the K$-$K' processes can be distinguished as they form a clear 3-fold symmetric pattern while the $\Gamma -$K processes capture 6-fold symmetry. We note that the nodal gap functions for $A_1$ and $A_2$ representations result in v-shaped DOS, see Fig.~\ref{fig:DOS}(d,e).
In Fig.~\ref{fig:QPI_triplet_z}(c) we show the QPI map for the $E$ representation. The gap function $\Delta_{E,1}^{{\rm t},z}$ is non-zero for the considered Fermi energy, and the global gap is opened, see Fig.~\ref{fig:DOS}(f). The Bogoliubov quasiparticles are gaped, however, the amplitude of the gap for the $\Gamma$ pocket is larger than for the K pockets. We note that the gap amplitude drops to zero at the K point, therefore the inner K pocket is less gapped than the outer one. Due to the finite broadening the K intravalley QPI contour is enhanced as well as the K$-$K' intervalley contours in comparison to the processes involving the $\Gamma$ pocket.

\begin{figure*}[t]
\centering
    \includegraphics[width=16cm]{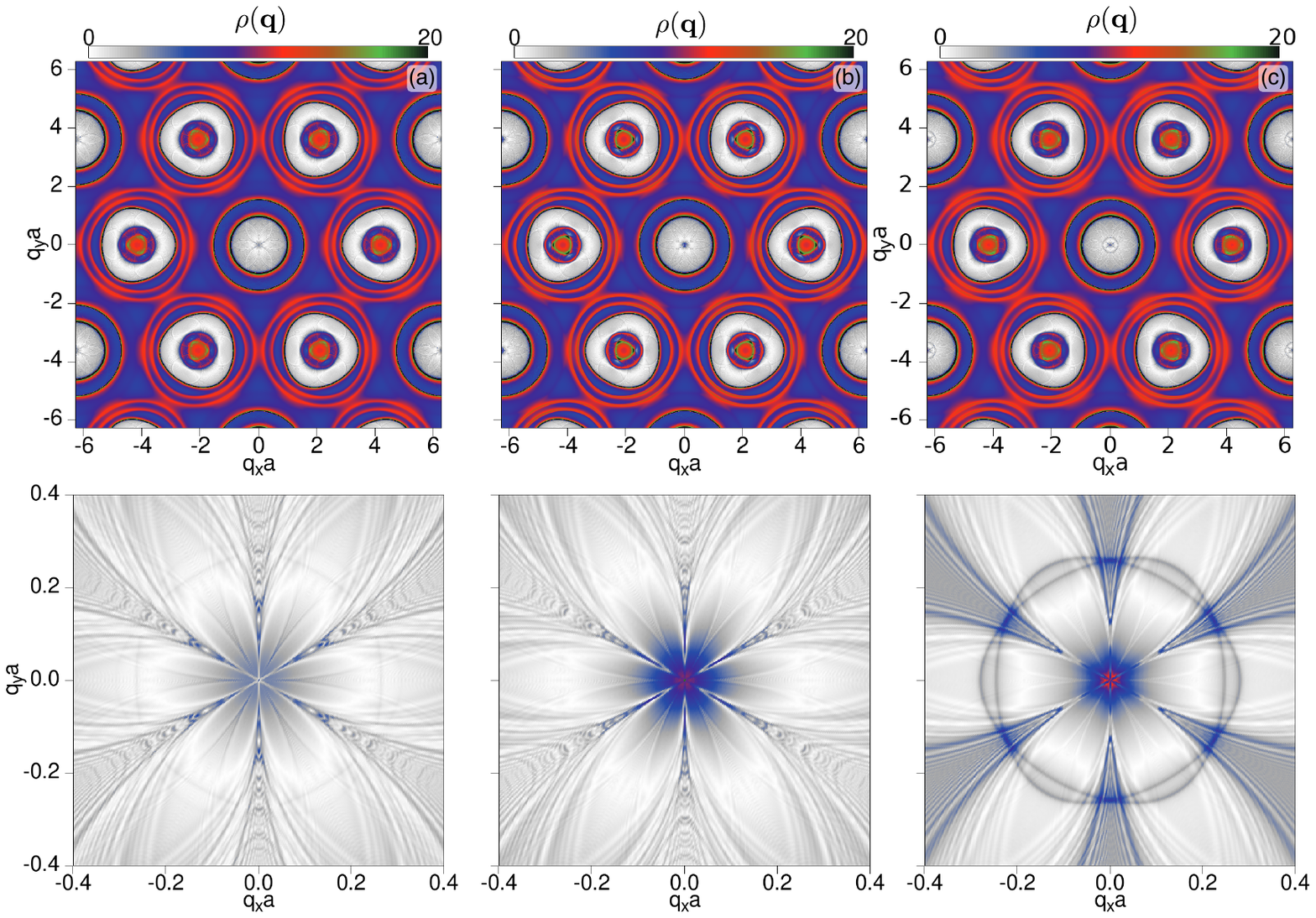} 
\caption{Calculated QPI patterns in the superconducting state for triplet order parameters with pseudovector $(d_x, d_y)$ multiplet for (a)~$A_1$, (b)~$A_2$, and (c)~$E$ representations. Plots in the bottom row are the zone center zooms for the corresponding top row.
The other parameters are as in Fig.~\ref{fig:QPI_singlet}.
}
\label{fig:QPI_triplet_xy}
\end{figure*}

Finally, we analyze the triplet pairing constructed using the multiplet $(d_x,d_y)$.
The QPI patterns for the $A_1$, $A_2$, and $E$ representations are similar, containing significant contour traces of the K intravalley and all the intervalley scattering processes except the $\Gamma$ intervalley scattering which is suppressed, see Fig.~\ref{fig:QPI_triplet_xy}. The suppression is due to the gap opened for the $\Gamma$ pocket for $A_1$ and $A_2$ representations, while for the $E$ representation the gap has a node along the $\Gamma$-K lines. In Fig.~\ref{fig:DOS}(g,h) we plot DOS for the $A_1$ and $A_2$ representations. The main coherence peaks enclose the gap opened for the $\Gamma$ pocket with a constant density corresponding to linearly dispersing in gap states around the K valleys forming the narrow v-shaped dependence near the zero energy. DOS for $E$ representation shows a broad finite valued v-shaped dependence, see Fig.~\ref{fig:DOS}, due to the gapless spectrum around K valleys and nodal features on the $\Gamma$ pocket along the $\Gamma$-K directions.
Nematicity of the singlet and triplet with $d_z$ gap functions for the two-dimensional representation $E$, as well as the chiral symmetry violation of the triplet with $(d_x,d_y)$ multiple, can be detected with QPI. We discuss it in Appendix \ref{sec:app_nematicity} and \ref{sec:app_PH_symm}.

\section{Conclusion}\label{Conclusion}
For a single-band model of NbSe$_2$ monolayer, Ising superconductor with Rashba spin-orbit coupling, we constructed using a group theoretical approach all possible superconducting gap functions beyond the nearest-neighbor approximation for all irreducible representations of ${\bf C}_{3{\rm v}}$ symmetry. We found the nodal gap function for the singlet pairing in $A_2$ representation and non-unitary triplet gap functions for two-dimensional $E$ representation.
Our analysis indicates that the nodal and nematic superconducting pairing, recently observed experimentally can be connected to the nodal and two-dimensional superconducting pairing functions.
Breaking the unitarity is associated with the asymmetrical band dispersion in the superconducting energy spectra.
Calculated QPI for single scalar impurity revealed characteristic patterns for the pairing functions.
The symmetry-based approach of combining group theoretical analysis and the QPI technique paves the way toward understanding exotic superconductivity.
We showed that QPI can capture nematic response for the two-dimensional representation that breaks time-reversal symmetry for singlet and triplet with $d_z$, and chiral symmetry violation for the triplet with $(d_x, d_y)$ multiplet.
Our approach to superconducting gap function construction can be extended to the case of an arbitrary class of the point group symmetry~\cite{T73} and the case of multiorbital superconductivity.
\acknowledgments
\textbf{J.H.}~acknowledges the financial support provided by the  Ministry of Education, Science, Research and Sport of the Slovak Republic.
\textbf{M.M.}~acknowledges the financial support provided by the Ministry of Education, Science, and Technological Development of the Republic of Serbia. This project has received funding from the European Union's Horizon 2020 Research and Innovation Programme under the Programme SASPRO 2 COFUND Marie Sklodowska-Curie grant agreement No.~945478.
\textbf{M.G.}~acknowledges financial support from Slovak Research and Development Agency provided under Contract No. APVV-20-0425, and Slovak Academy of Sciences project IMPULZ IM-2021-42 and project FLAG ERA JTC 2021 2DSOTECH.
\appendix

\section{Model Hamitonian}\label{sec:DFT-TB}

\subsection{Effective tight-binding model for {NbSe$_{2}$}}
To describe efficiently electronic states of NbSe$_{2}$ monolayer we consider an effective single-orbital tight-binding model Hamiltonian \cite{SM19}.
The orbital part of the Bloch Hamiltonian up to the seventh Nb neighbor atoms reads

\begin{eqnarray}
&&{\cal H}_{\rm orb}({\bf k})=\varepsilon_0 + 2 t_1 (\cos{2\alpha}+2\cos{\alpha}\cos{\beta})+
2 t_2 (\cos{2\beta}\nonumber\\
+&&2\cos{3\alpha}\cos{\beta})+2t_3 (\cos{4\alpha}+2\cos{2\alpha}\cos{2\beta})\nonumber\\
+&&4t_4(\cos{\alpha}\cos{3\beta}+\cos{4\alpha}\cos{2\beta}+\cos{5\alpha}\cos{\beta})+2t_5\times\nonumber\\ &&(\cos{6\alpha}+2\cos{3\alpha}\cos{3\beta})+2 t_6 (\cos{2\beta}+2\cos{3\alpha}\cos{\beta})\nonumber\\
+&&4t_7(\cos{7\alpha}\cos{\beta}+\cos{5\alpha}\cos{3\beta} +
\cos{2\alpha}\cos{4\beta}),
\end{eqnarray}

where $\alpha=k_x a/2$, $\beta=\sqrt{3}k_y a/2$, $a$ is the lattice constant, $k_x$ and $k_y$ are components of the wave vectors in Cartesian frame, $\varepsilon_0$ is the energy offset of the corresponding to the chemical potential of the electron-doped NbSe$_2$ layer with respect to the isolated NbSe$_2$ layer, and $t_i$, $i=1,...,7$ are the real hopping parameters.
The model describes Nb $d$-band close to the Fermi level.
By including intrinsic SOC in accordance with the ${\bf D}_{3{\rm h}}$ point group of the free-standing NbSe$_2$ monolayer, the effective $d$-band split and the horizontal mirror plane $\sigma_{\rm h}$ constrains spins in the out-of-plane direction.
The intrinsic spin-orbit coupling Hamiltonian $H_{\rm I}$~\cite{SM19} up to the third neighbor equals to
\begin{eqnarray}
{\cal H}_{\rm I}({\bf k})=2\sigma_z\Big(&\lambda_{\rm I}^{(1)}&
(\sin{2\alpha}-2\sin{\alpha}\cos{\beta})\nonumber\\
+&\lambda_{\rm I}^{(3)}&(\sin{4\alpha}-2\sin{2\alpha}\cos{2\beta})\Big)\,,
\end{eqnarray}
where $\sigma_z$ is the Pauli matrix, and the nonzero real parameters $\lambda_{\rm I}^{(1)}$ and $\lambda_{\rm I}^{(3)}$ quantify the strength of the interaction in the first and third neighbor approximation. The parameter $\lambda_{\rm I}^{(2)}=0$ due to vertical mirror plane symmetry.

When a monolayer of NbSe$_2$ is placed on a substrate, the horizontal mirror plane symmetry $\sigma_{\rm h}$ is broken, triggering the appearance of Rashba spin-orbit due to a proximity effect by virtue of an electrical field perpendicular to the plane. This leads to a reduction of the ${\bf D}_{3{\rm h}}$ symmetry to ${\bf C}_{3{\rm v}}$. The effective Rashba spin-orbit coupling can be derived similarly as the intrinsic counterpart. In real space, the Rashba Hamiltonian has the following form

\begin{equation}\label{RashbaSOC}
    H_{\rm R}={\rm i}\sum_{l,i}\lambda_{\rm R}^{(l)}({\bf e}_z\times {\bf e}_i^{(l)})\cdot{\bm \sigma},
\end{equation}

where $\lambda_{\rm R}^{(l)}$ are the real-valued Rashba parameters for the $l$-th order, ${\bf e}_z$ is the unit vector in the $z$-direction, and ${\bf e}_i^{(l)}={\bf d}_i^{(l)}/|{\bf d}_i^{(l)}|$ is the normalized distance of each Nb atom from the centered one.
By performing the Fourier transform of the Rashba Hamiltonian, we obtain the following form in the $k$-space
\begin{equation}
    {\cal H}_{\rm R}({\bf k})={\rm i}\sum_{l,i}\lambda_{\rm R}^{(l)}
    {\rm e}^{{\rm i}{\bf k}\cdot{\bf d}_i^{(l)}}({\bf e}_z\times {\bf e}_i^{(l)})\cdot{\bm \sigma}.
\end{equation}
Next, we consider contributions up to the third neighbors, obtaining \begin{eqnarray}
&&{\cal H}_{\rm R}({\bf k})=2\lambda_{\rm R}^{(1)}\Big(\sqrt{3}\cos{\alpha}\sin{\beta}\sigma_x
     -f(\alpha,\beta)\sigma_y\Big)\nonumber\\
&&+2\lambda_{\rm R}^{(2)}\Big((\sin{2\beta}+\cos{3\alpha}\sin{\beta})
\sigma_x-\sqrt{3}\sin{3\alpha}\cos{\beta}\sigma_y
    \Big)\nonumber\\
&&+2\lambda_{\rm R}^{(3)}\Big(\sqrt{3}\cos{2\alpha}\sin{2\beta}\sigma_x-
     f(2\alpha,2\beta)\sigma_y
    \Big).
\end{eqnarray}
where $f(\alpha,\beta)=(\sin{2\alpha}+\sin{\alpha}\cos{\beta})$, $\sigma_x$ and $\sigma_y$ are Pauli matrices, while 
$\lambda_{\rm R}^{(1)}$, $\lambda_{\rm R}^{(2)}$, $\lambda_{\rm R}^{(3)}$
are parameters to be determined by fitting the DFT data to the model Hamiltonian 
${\cal H}_{\rm orb}({\bf k})+{\cal H}_{\rm I}({\bf k})+{\cal H}_{\rm R}({\bf k})$. In addition to the Rashba parameters, the orbital $\varepsilon_0$, $t_i$, $i=1,...,7$ and intrinsic $\lambda_{\rm I}^{(1)/(3)}$ parameters need to be determined. 

\begin{table}[htp]
\caption{Fitting parameters of the single-orbital band model of NbSe$_2$. Besides the energy offset $\varepsilon_0$, seven hopping parameters $t_i$, $i=1,...,7$, model the spin-independent band structure. Spin-orbit properties are described using the parameters $\lambda_{\rm I}^{(1)/(2)}$ and $\lambda_{\rm R}^{(1)/(2)/(3)}$, representing the interaction strengths of the intrinsic and Rashba spin-orbit Hamiltonian.}\label{tab:hoppings}
\centering
\begin{tabular}{cc}\hline\hline
$\varepsilon_0$\,[meV]& -346.26\\ \hline
$t_1$\,[meV]& 33.52\\ 
$t_2$\,[meV]& 97.26\\ 
$t_3$\,[meV]&-2.11\\ 
$t_4$\,[meV]&-13.53\\ 
$t_5$\,[meV]&-10.30\\ 
$t_6$\,[meV]&3.48\\ 
$t_7$\,[meV]&1.69\\ \hline
$\lambda_1$\,[meV]    & 13.27 \\
$\lambda_2$\,[meV]    & -1.94 \\\hline
$\lambda_{\rm R}^1\,[\mu {\rm eV}]$    &  -9.60 \\
$\lambda_{\rm R}^2\,[\mu {\rm eV}]$    &  -0.29  \\
$\lambda_{\rm R}^3\,[\mu {\rm eV}]$    &  -4.70 \\ \hline\hline
\end{tabular}
\end{table}

\subsection{Fitting the model to the DFT data}
To obtain relevant parameters for the single-band model, which is, besides the superconducting gap function, one of the inputs in BdG Hamiltonian,
we performed DFT calculations of electronic structure for the NbSe$_2$ monolayer. This was done using the plane wave DFT suite QUANTUM ESPRESSO (QE) package~\cite{QE1,QE2} using the full relativistic SG15 Optimized Norm-Conserving Vanderbilt (ONCV) pseudopotentials~\cite{H13,SG15,SGH16}, with the kinetic energy cut-offs for the wave function and charge density 45\,Ry and 180\,Ry, respectively.  
For the Brillouin zone sampling, a $12\times 12\times1$ $k$-points mesh was considered using the Monkhorst-Pack scheme. The energy convergence threshold for self-consistent calculation, including the spin-orbit coupling, was sent to $10^{-10}$ Ry/bohr. 
A vacuum of 15 ${\rm \AA}$ in the $z$-direction was used.
\begin{figure}[h]
\centering
\includegraphics[width=8.4cm]{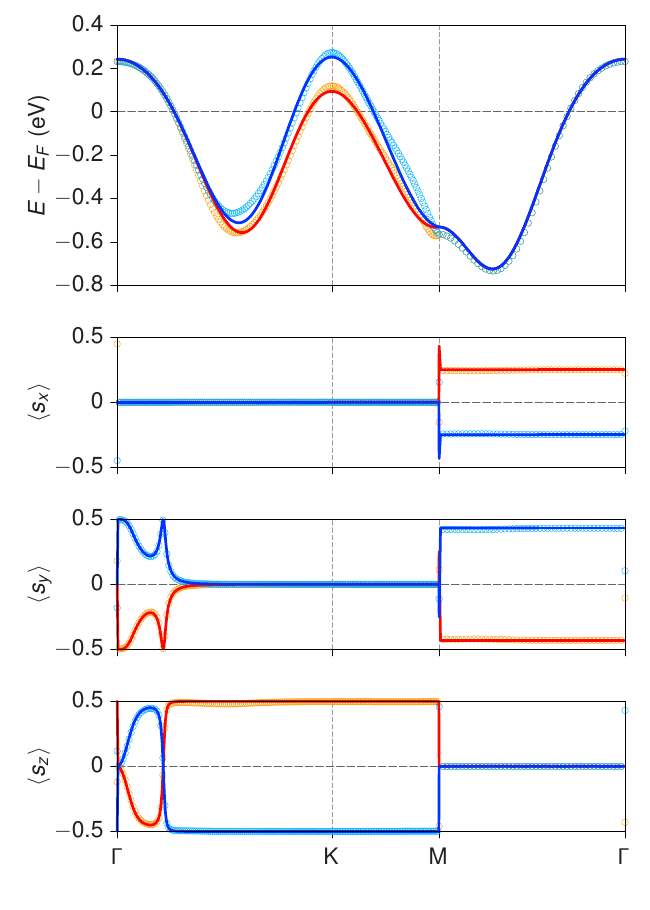}
\caption{Calculated energy bands dispersion close to the Fermi level along high symmetry lines in the first Brillouin zone for NbSe$_2$ monolayer in the electric field of 0.1~V/nm perpendicular to the monolayer plane, and the corresponding spin expectation values. The circles are DFT date, and the solid lines are effective single orbital tight-binding model.}\label{fig:model}
\end{figure}
We obtained the orbital and spin-orbital hopping parameters by fitting the DFT data for the NbSe$_2$ monolayer in a perpendicular field of 0.1~V/nm. The parameters are gathered in Table~\ref{tab:hoppings}. A comparison between the numerical band structure and spin expectation values and our model is shown in Fig.~\ref{fig:model}.
\section{Density of states}\label{sec:DOS}
\begin{figure*}[t]
\centering
\includegraphics[width=16cm]{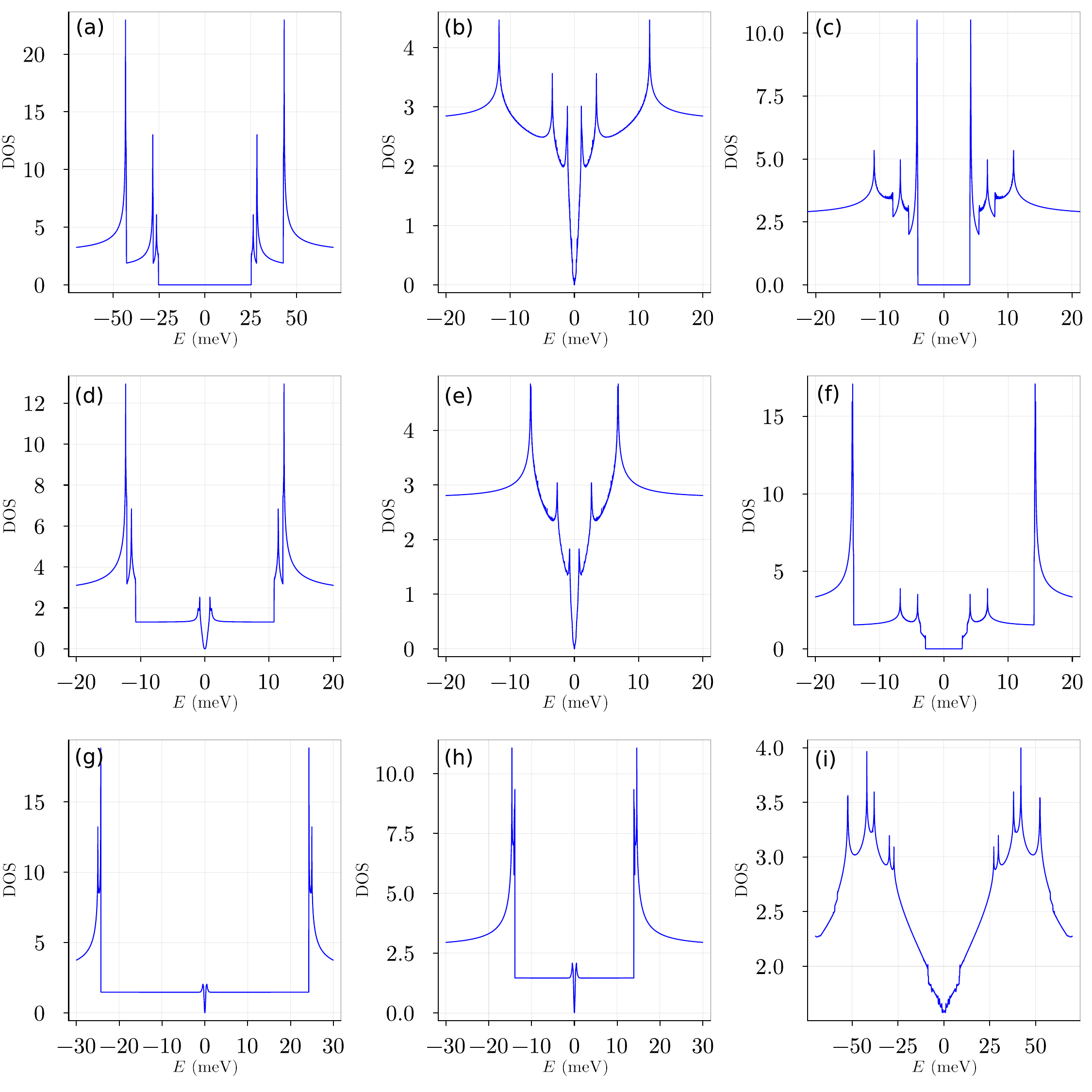}
\caption{Calculated density of states for NbSe$_2$ monolayer in the electric field of 0.1~V/nm perpendicular to the monolayer plane for singlet gap functions (a)~for $A_1$ representation; (b)~for $A_2$ representation; (c)~for $E$ representation. Density of states for triplet gap functions with pseudovector component $d_z$ (d)~for $A_1$ representation; (e)~$A_2$ representation; (f)~for $E$ representation; and for pseudovector multiplet $(d_x, d_y)$ (g)~$A_1$ representation; (h)~for $A_2$ representation; (i)~for $E$ representation. For calculations the amplitude $\Delta_0 = 10$~meV was used.}
\label{fig:DOS}
\end{figure*}

In Fig.~\ref{fig:DOS} we plot the density of states (DOS) 
as a function of the quasiparticle energy for the BdG Hamiltonian considering superconducting gap functions derived in Sec.~\ref{Sec:Gaps}. The energy dependencies were calculated using the linear tetrahedron method with $8192 \times 8192$ $k$-point sampling of the first BZ. We consider $\Delta_0 = 10$~meV in order to reduce the computational complexity in terms of $k$-sampling. We note that the overall qualitative dependencies of the DOS are unaffected when compared to $\Delta_0 = 1$~meV. For the singlet pairing the gap has finite value for $A_1$ and $E$ representations, see Fig.~\ref{fig:DOS}(a,c). The K pocket states form the gap coherence peaks for the $A_1$ representation, while for the $E$ the peaks originate from the $\Gamma$ pocket. For the $A_2$ representation the nodal line character of the gap function results in a v-shape dependence, see Fig.~\ref{fig:DOS}(b).

For the triplet pairing with $d_z$ pseudovector component the gap functions for $A_1$ and $A_2$ representations show the v-shape gap, see Fig.~\ref{fig:DOS}(d,e). For the $E$ representation the global gap is open with coherence peaks originating from the K pocket.

The triplet pairing with $(d_x, d_y)$ multiplet shows similar v-shaped DOS for the gap functions for the $A_1$ and $A_2$, see Fig.~\ref{fig:DOS}(g,h). In the case of the $E$ representation, see Fig.~\ref{fig:DOS}(i), the DOS is finite with a rather broad v-shape dependence around the zero energy.

\section{Nematicity response in QPI}\label{sec:app_nematicity}
\begin{figure*}[t]
\centering
\includegraphics[width=15cm]{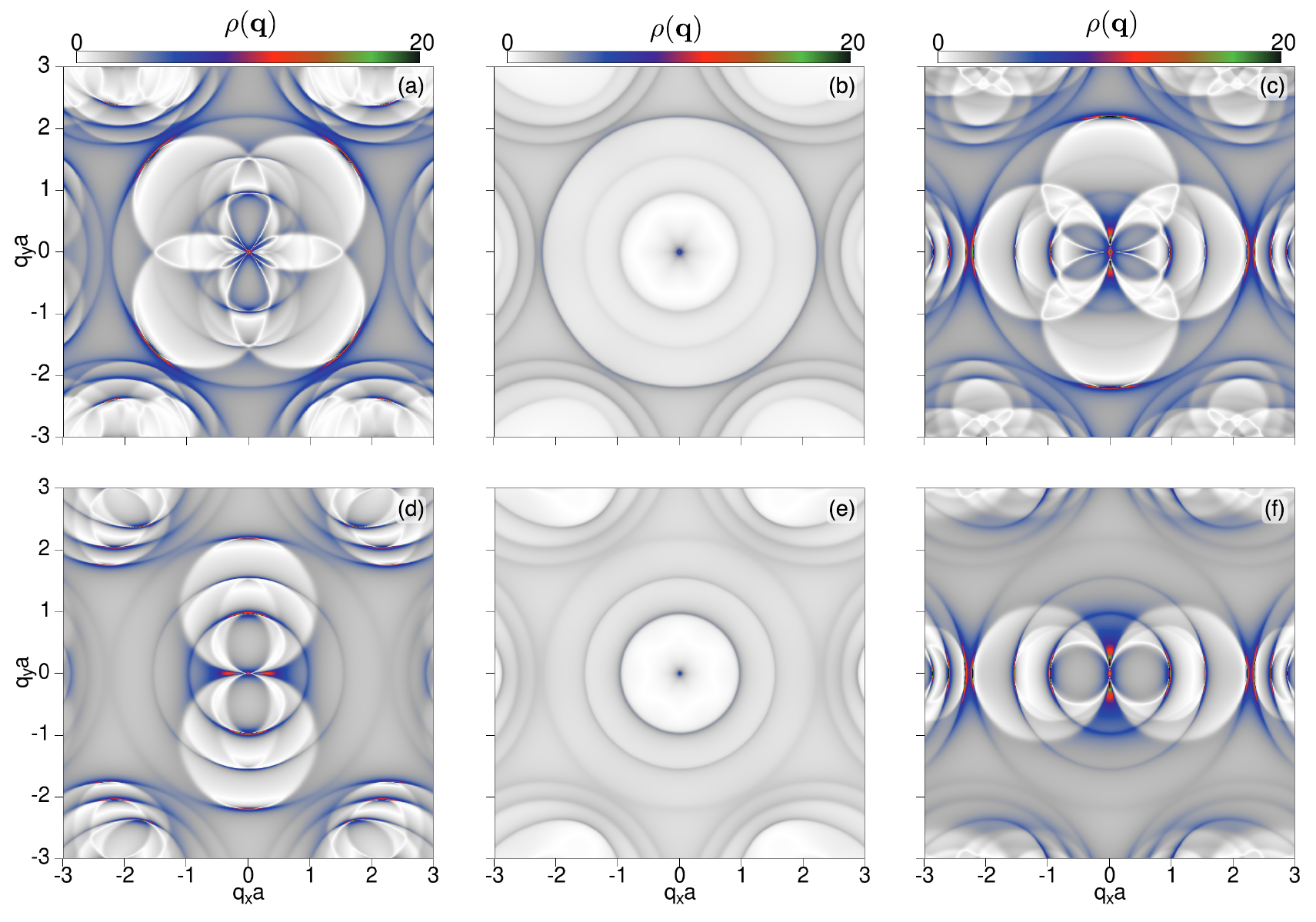}
\caption{Calculated QPI patterns in the superconducting state for $E$ representation with $(d_x, d_y)$ multiplet for different phase angles. 
Singlet pairing (a)~for $\varphi=\pi/4$, (b)~$\varphi=\pi/2$, (c)~$\varphi=3\pi/4$. 
Triplet pairing with pseudovector $d_z$ (d)~for $\varphi=\pi/4$, (e)~$\varphi=\pi/2$, (f)~$\varphi=3\pi/4$. The other parameters are as in Fig.~\ref{fig:QPI_singlet}.
}
\label{fig:QPI_nematic}
\end{figure*}

The gap functions for the two-dimensional representation $E$ in singlet and triplet pairing lead to a nematic phase with a global gap. This pairing type belongs to the chiral topological class \cite{Chiu2016:RMP}. Any combination of the form $\Delta_E^{\mathcal{P}}(\varphi) = \cos(\varphi)\Delta_{E,1}^{\mathcal{P}} + \sin(\varphi)\Delta_{E,2}^{\mathcal{P}}$ follows the irreducible representation and can be detected by the QPI. The gap functions in the lowest order in momentum in the vicinity of the Brillouin zone center are listed in Table~\ref{tab:nematic_gaps_E}. The $\varphi$ rotates the gap functions where for the singlet pairing one recognizes the rotated $d+\mathrm{i}d$ gap, while for the triplet with $d_z$ pseudovector, it rotates the $p + {\rm i} p$ type gap.
In Fig.~\ref{fig:QPI_nematic} we plot the QPI patterns for the singlet and triplet with $d_z$, $\mathcal{P}=\{s, ({\rm t},z)\}$ and phase angle $\varphi = \{\pi/4, \pi/2, 3\pi/4\}$. In case of $\Delta_E^{{\rm t},z}(\pi/4)$ the QPI pattern resembles $p_y$-like orbital for the small $\mathbf{q}$. This can be understood as the gap function having reduced amplitude for $k_x \simeq 0$ allowing due to the finite lifetime of quasiparticles an enhanced scattering along the $q_y$. Similarly one explains other calculated QPI patterns.  

\begin{table}[!h]
    \centering
    \begin{tabular}{@{\quad}c@{\quad}|@{\quad}c@{\quad}@{\quad}c@{\quad}@{\quad}c@{\quad}}
    \hline\hline
      $\varphi$                            & $\pi/4$          & $\pi/2$           & $3\pi/4$ \\ \hline
      $\Delta_E^{\rm s}(\varphi)/d_0$      & $k_x^2 - k_y^2$  & $(k_x - i k_y)^2$ & $i k_x k_y$ \\
      $\Delta_E^{{\rm t},z}(\varphi)/d_z$ & $k_x$            & $(k_x - i k_y)$   & $i k_y$ \\
      \hline\hline
    \end{tabular}
    \caption{Nematic gap functions dependencies near the Brillouin zone center for different phase angles.}
    \label{tab:nematic_gaps_E}
\end{table}

\section{Chiral symmetry response in QPI}\label{sec:app_PH_symm}
\begin{figure}[t]
    \centering
    \includegraphics[width=5.6cm]{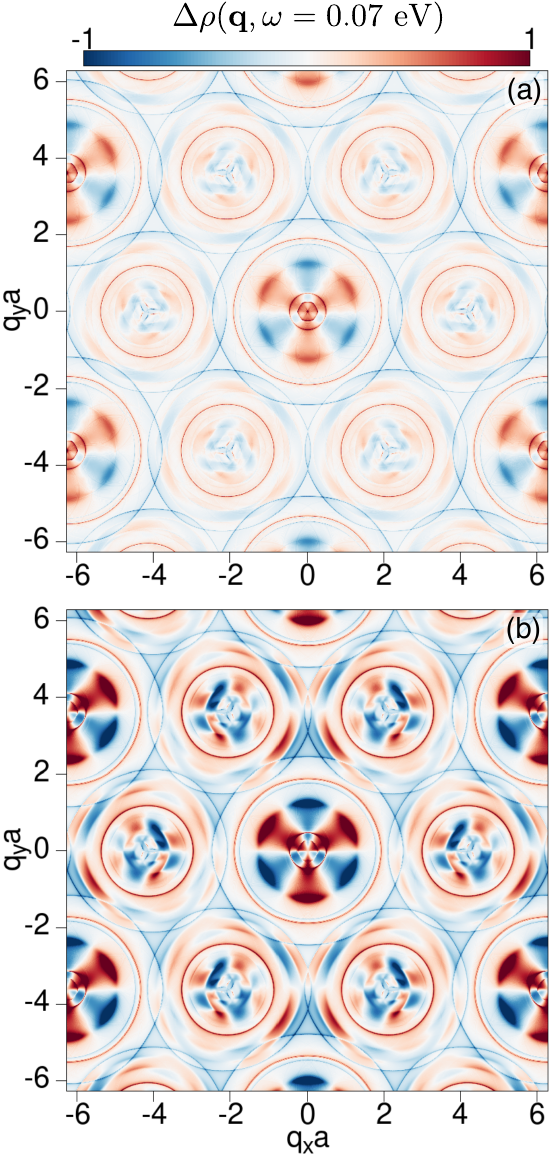}
    \caption{Calculated difference of the QPI patterns $\Delta \rho(\mathbf{q},\omega) = \rho(\mathbf{q},\omega) - \rho(\mathbf{q},-\omega)$ in the superconducting triplet state for $E$ representation with pseudovector $(d_x, d_y)$ multiplet for $\Delta_0 = 10$~meV and $\omega = 0.07$~eV.
    (a)~phase angle $\varphi = 0$; and
    (b)~$\varphi = \pi/3$.
    }
    \label{fig:QPI_diff_E_t_xy}
\end{figure}

As discussed in Sec.~\ref{sec:triplet_dx_dy_gaps} the triplet $\Delta_E^{{\rm t},xy}$ gap function is non-unitary except for the special phase angles $\varphi$.
In Table~\ref{tab:E_t_xy_symm} we list the symmetry properties of the $\Delta_E^{{\rm t},xy}$ gap function.
The ${\bf C}_{\rm 3v}$ symmetry is reduced to ${\bf C}_1$ for $\varphi\in 3\pi/4 + n\pi$ and to ${\bf C}_{\rm 1v}$ for $\varphi\in \pi/4 + n\pi$, where $n$ is the integer.
Along with the unitarity of the $\Delta_E^{{\rm t},xy}$ gap function, the chiral symmetry of the BdG quasiparticle spectra is recovered for the above special angles.
In Fig.~\ref{fig:QPI_diff_E_t_xy} we plot differences of the QPI signals $\Delta \rho(\mathbf{q},\omega) = \rho(\mathbf{q},\omega) - \rho(\mathbf{q},-\omega)$ for $\Delta_E^{{\rm t},xy}$ and energy $\omega=0.07$~eV that falls to the spectral gap opened between electron and hole spin down quasiparticle bands near the K points. The non-zero signal of the $\Delta \rho(\mathbf{q},\omega)$ demonstrates the chiral symmetry violation of the BdG quasiparticle spectra. For the case of $\varphi = 0$, Fig.~\ref{fig:QPI_diff_E_t_xy}(a), the signal preserves the threefold symmetry, while for the $\varphi = \pi/3$, Fig.~\ref{fig:QPI_diff_E_t_xy}(b), the pattern breaks the threefold symmetry. 

\begin{table}[!h]
    \centering
    \begin{tabular}{l|@{\quad}c@{\quad}|@{\quad}c@{\quad}}
    \hline\hline
      $\Delta^{{\rm t},xy}_E$ & unitary & chiral\\ \hline
    ${\bf C}_{\rm 3v} \xrightarrow{\varphi} {\bf C}_1$ & 0 & 0 \\
    ${\bf C}_{\rm 3v} \xrightarrow{\varphi\in (3\pi/4 + n\pi )} {\bf C}_{1}$ & 1 & 1 \\
    ${\bf C}_{\rm 3v} \xrightarrow{\varphi\in (\pi/4 + n\pi )} {\bf C}_{1{\rm v}}$ & 1 & 1 \\
    \hline\hline
    \end{tabular}
\caption{Symmetries of the $\Delta_E^{{\rm t},xy}$ gap function for different phase angles $\varphi$.}
\label{tab:E_t_xy_symm}
\end{table}

\bibliography{bibliography}
\end{document}